\pdfoutput=1  
\documentclass[aps,prl,superscriptaddress,floatfix,nofootinbib,preprintnumbers,eqsecnum,twocolumn,dvipsnames]{revtex4-1}
\counterwithout{equation}{section}
\usepackage{amsmath,float,graphicx,placeins,amssymb,multirow,pgfplots,pgfplotstable}
\usepackage{empheq}
\usepackage{tikz}
\usepackage{comment}
\usepackage{enumerate,slashed,cancel,comment,color,bbm,graphicx,rotating,placeins,mathtools,multirow,hhline}
\usepackage[normalem]{ulem}
\usepackage{array}

\usepackage{hyperref}
\usepackage{color}
 \hypersetup
 {
   colorlinks,
   linkcolor={blue!80!black},
   citecolor={blue!70!black},
   urlcolor={blue!70!black}
 }

\usepackage{lipsum}
\usepackage{diagbox}
\usetikzlibrary{shapes.geometric,arrows}
\graphicspath{}

\def\beq{\begin{equation}}
\def\eeq{\end{equation}}
\def\beqn{\begin{eqnarray}}
\def\eeqn{\end{eqnarray}}

\def\be{\begin{equation}}
\def\ee{\end{equation}}
\def\bea{\begin{eqnarray}}
\def\eea{\end{eqnarray}}

\begin{document}

\title{Primordial Black Holes Are True Vacuum Nurseries}


\preprint{IPPP/23/65}
\author{Louis Hamaide}
 \email[Email (Corresponding Author): ]{l.hamaide@ucl.ac.uk}
 \affiliation{Theoretical Particle Physics and Cosmology, King’s College London,\\ Strand, London WC2R 2LS, United Kingdom}
  \affiliation{AMOPP, University College London,\\ Gower Place, London WC1E 6BS, United Kingdom}
  \author{Lucien Heurtier}
\email[Email: ]{lucien.heurtier@durham.ac.uk}
 \affiliation{Theoretical Particle Physics and Cosmology, King’s College London,\\ Strand, London WC2R 2LS, United Kingdom}
\affiliation{IPPP, Durham University, Durham, DH1 3LE, United Kingdom}
\author{Shi-Qian Hu}
\email[Email: ]{shiqian.hu@kcl.ac.uk}
 \affiliation{Theoretical Particle Physics and Cosmology, King’s College London,\\ Strand, London WC2R 2LS, United Kingdom}
\author{Andrew Cheek}
 \email[Email: ]{acheek@camk.edu.pl}
 \affiliation{Astrocent, Nicolaus Copernicus Astronomical Center of the Polish Academy of Sciences, ul.Rektorska 4, 00-614 Warsaw, Poland}

\begin{abstract} 
The Hawking evaporation of primordial black holes (PBH) reheats the Universe locally, forming hot spots that survive throughout their lifetime. We propose to use the temperature profile of such hot spots to calculate the decay rate of metastable vacua in cosmology, avoiding inconsistencies inherent to the Hartle-Hawking or Unruh vacuum. We apply our formalism to the case of the electroweak vacuum stability and find that a PBH energy fraction $\beta > 7\times 10^{-80} (M/\mathrm{g})^{3/2}$ is ruled out 
for black holes with masses $0.8 \,{\rm g} < M < 10^{15} \,{\rm g}$.
\end{abstract}
\maketitle


\def\beq{\begin{equation}}
\def\eeq{\end{equation}}
\def\beqn{\begin{eqnarray}}
\def\eeqn{\end{eqnarray}}

\paragraph{Introduction. ---\label{sec:intro}}
Many models of early-universe cosmology predict the formation of light primordial black holes that have evaporated long before the onset of Big Bang Nucleosynthesis (BBN). Although such evanescent PBHs constitute unique relics of primordial cosmic history, confirming their existence with observations can be challenging, if not impossible, due to their small density fraction in cosmology.

Hawking's famous result states that light black holes can be extremely hot, $T_H = (8\pi G M)^{-1}$, which would have substantial impact on the local environment. By using the so-called Hartle-Hawking vacuum --- in which PBHs are assumed to be in thermal equilibrium with their surrounding plasma~\cite{Hartle:1976tp} --- the authors of Refs.~\cite{Gregory:2013hja, Burda:2015isa, Burda:2015yfa, Burda:2016mou} proposed that PBHs could trigger first-order phase transitions in the early Universe and endanger the meta-stability of the electroweak vacuum. Indeed, unlike sphaleron processes which pump energy directly from a plasma with infinite heat capacity to form a true vacuum bubble, PBHs have the ability to release part of their mass energy to support the formation of true-vacuum bubbles around them. If true, such a claim would severely constrain the formation of PBHs with masses smaller than $\mathcal O(10)\mathrm{g}$ in the early Universe or impress the importance of stabilising the Higg's vacuum via new physics~\cite{Hiller:2022rla}. 

Since then, this claim has been subject to controversy: PBHs are rarely surrounded by a plasma of temperature $T=T_H$ and several authors pointed out that the Unruh vacuum \cite{Unruh:1976db} --- in which PBHs radiate energy in an empty Universe --- should be used instead, largely mitigating the aforementioned result~\cite{Kohri:2017ybt, Hayashi:2020ocn, Shkerin:2021zbf, Shkerin:2021rhy, Strumia:2022jil, Briaud:2022few}. Moreover, thermal quantum corrections to the Higgs potential were argued to be relevant for PBHs lighter than $\mathcal O(10^{3})\mathrm{g}$, increasing the energy of true vacuum bubbles, and effectively rescuing the stability of the electroweak vacuum~\cite{Strumia:2022jil}. 

In this {\em letter}, we point out that evaporating PBHs neither live in the vacuum nor are in thermal equilibrium with their surrounding. Instead, by depositing energy locally into the thermal plasma, Hawking radiation supports the formation of hot spots around PBHs~\cite{Das:2021wei,He:2022wwy}. Building on the results derived in~\cite{Gregory:2013hja, Burda:2015isa, Burda:2015yfa, Burda:2016mou, Dai:2019eei}, we provide a new avenue to calculate false-vacuum decay (FVD) rates around PBHs in cosmology. In particular, we will show that the presence of such PBH hot spots can indeed seed the formation of true  electroweak vacuum bubbles in the early Universe, circumventing the criticisms listed above. 

%

\paragraph{Bubble Action at $T<T_H$. ---}
Let us for the time being consider an homogeneous scalar field configuration $\phi$ living in a metastable minimum of its potential $V(\phi)$. The decay rate of this configuration is determined in curved spacetime by a saddle point ``bounce" solution of the Euclidean action. Around a Schwarzschild black hole living in a background plasma with an arbitrary constant temperature $T$, one can write the Euclidean action of a time-independent scalar field bubble configuration as in \cite{Burda:2016mou, Dai:2019eei}
\be\label{eq:action}
I_{\rm b}[T] = \beta\int dx^3 \, \sqrt{-h}\left(-\frac{R}{16\pi G}+\frac{1}{2}h^{\mu\nu}\partial_\mu \phi\partial_\nu \phi + V(\phi)\right)\,.
\ee
in which $h_{\mu\nu}$ is the spatial part of the metric --- assumed to be time-independent as well ---
and $\beta=1/T$ denotes the Euclidean periodicity. At first sight, because the periodicity $\beta$ appears in this expression as an overall factor, it is tempting to conclude that a low 
plasma temperature would lead to a large bubble action, hence suppressing the false-vacuum decay rate exponentially:
\be
T\ll T_H \ \Rightarrow\  \Gamma_{\rm FVD}\propto\exp\left(-I_{\rm b}[T]\right)\ll \exp\left(-I_{\rm b}[T_H]\right)\,.
\ee
However, it is important to note that there are two pieces missing in Eq.~\eqref{eq:action}~\cite{Gregory:2013hja}. First, the gravitational background action should be subtracted to extract the actual energy of the bubble. Second, if the temperature entering the Euclidean action's periodicity differs from the Hawking temperature at the black hole horizon, the system features a conical deficit at the horizon that needs to be accounted for in the calculation. Ref.~\cite{Gregory:2013hja} calculated both contributions in the thin-wall approximation. In particular, the contribution from the conical deficit was evaluated using a mathematical regularisation procedure: using a temperature profile interpolating over a scale $\epsilon$ between a constant temperature $T(r)=T$ for $r-r_H\gg \epsilon$, and $T(r_H)=T_H$ at the horizon, one can show that the Ricci contribution to the action is of order
\be\label{eq:conical}
I_{\rm b}[T] \supset -\frac{\beta_H - \beta}{\beta}\frac{\mathcal A}{4G}+\mathcal O(\epsilon)\,,
\ee 
in the $\epsilon\to 0$ limit. In this equation $\mathcal A$ stands for the area of the black hole, and $\beta_H = 1/T_H$ the corresponding inverse Hawking temperature.

Remarkably, after summing up all the contributions, the authors of Ref.~\cite{Gregory:2013hja} proved, in the $\epsilon\to 0$ limit, that the $\beta$-dependency of the total Euclidean action exactly cancels, leading to the seminal result
\be\label{eq:RuthResult}
I_{\rm b}[T] = \frac{\mathcal A_+}{4G}-\frac{\mathcal A_-}{4G} = I_{\rm b}[T_H]\,,
\ee
where $\mathcal A_+$ ($\mathcal A_-$) denotes the area of the black hole horizon before (after) the formation of the bubble. This property was then generalised beyond the thin-wall approximation, and in the presence of matter in Ref.~\cite{Burda:2015yfa}. Despite its remarkable generality, to our knowledge, this result was only used in the Hartle-Hawking vacuum where $T=T_H$ to derive constraints on PBHs using the electroweak vacuum stability. In what follows, we will show that in a realistic cosmological set-up, PBHs do live in a plasma with a temperature lower than their Hawking temperature, but also that the mathematical smoothing used to derive Eq.~\eqref{eq:conical} does actually correspond to a physical situation that we will describe.

\bigskip

\paragraph{Thermal Profile Around PBHs. ---}

Throughout cosmic history, PBHs are surrounded by two main sources of energy: $(i)$ their Hawking radiation, made of relativistic particles with average energies $\langle E\rangle \sim T_H$, and $(ii)$ the energy of the ambient plasma surrounding the black hole, with particles with average energy $\langle E\rangle \sim T$.
Studies considering the Hartle-Hawking vacuum assume that $T=T_H$ and both energy sources are in perfect equilibrium. Instead, studies working with the Unruh vacuum neglect the presence of any ambient plasma and only consider the Hawking radiation, effectively working in the limit where $T\to 0$.

As a matter of fact, both of these pictures are incomplete. When PBHs form, the Universe's temperature may be much larger than their Hawking temperature. However, the end of PBH evaporation typically takes place much later, when the Universe's temperature is much colder than its original Hawking temperature. Nevertheless, a big piece is missing in this discussion: 
In the intermediate period, the Standard Model (SM) radiation that forms Hawking radiation unavoidably interacts with the surrounding plasma. This energy deposition is localised in space, and instead of reheating the Universe in a homogeneous way, PBHs are expected to heat up their surrounding environments. 
This energy deposition leads to an inhomogeneous temperature profile forming around the PBHs that persists long after they have completely evaporated. As a consequence, the plasma temperature around PBHs can differ by orders of magnitude from the average temperature in the Universe.
 
The temperature profile evolution around PBHs was explored in Ref.~\cite{He:2022wwy, Das:2021wei}. Assuming a universal  coupling constant $\alpha$ to encode particle interactions between Hawking radiation and the plasma, the authors of Ref.~\cite{He:2022wwy} showed that for PBHs with masses 
\bea
M&\gtrsim& M_\star\,,\nonumber\\
M_\star &\equiv & 0.8\mathrm{g}\left(\frac{\alpha}{0.1}\right)^{-\frac{11}{3}}\,,
\eea
 the formation of a hot spot is quicker than the evaporation of the black hole. PBHs are thus rapidly surrounded by an initial hot spot featuring a constant temperature
 \bea\label{eq:plateau}
 T_{\rm plateau} &\approx& 2\times 10^{-4}\left(\frac{\alpha}{0.1}\right)^{\frac{8}{3}}T_H \nonumber\\
 &\times&\left(\frac{g_\star(T_{\rm plateau})}{106.75}\right)^{-\frac{2}{3}}\left(\frac{g_{\star}(T_H)}{106.75}\right)^{\frac{2}{3}}\,,
 \eea
 over a distance
 \bea
 r_{\rm plateau}&\approx & 7\times 10^{8}  \left(\frac{\alpha}{0.1}\right)^{-6}r_H\nonumber\\
 &\times &\left(\frac{g_\star(T_{\rm plateau})}{106.75}\right)\left(\frac{g_{\star}(T_H)}{106.75}\right)^{-1}\,,
 \eea
where 
$g_\star(T)$ denotes the number of degrees of freedom that are relativistic at temperature $T$. At a time smaller than the evaporation time $t_{\rm ev}\equiv \Gamma_{\rm ev}^{-1}$, where \mbox{$\Gamma_{\rm ev}\equiv |\Dot{M}/M|\propto M^{-3}$} can be calculated from e.g. Refs.~\cite{MacGibbon:1990zk, MacGibbon:1991tj}, the value of this plateau temperature is constant. On larger distances, diffusion dominates over energy deposition and the temperature decreases, smoothly interpolating between the plateau temperature and the Universe's temperature. 

Later on, the PBH mass starts decreasing, its Hawking temperature increases, and so does the plateau temperature, while its radius decreases. After the black hole mass reaches $M=M_\star$, the plateau temperature saturates at a maximum value~\cite{He:2022wwy}
\bea\label{eq:max}
T_{\rm max}&\approx& 2\times 10^9 \mathrm{GeV} \left(\frac{\alpha}{0.1}\right)^{\frac{19}{3}}\nonumber\\
&\times &\left(\frac{g_\star(T_{\rm max})}{106.75}\right)^{-\frac{4}{3}}\left(\frac{g_{\star}(T_H)}{106.75}\right)^{\frac{5}{6}}\,,
\eea
over a radius
\be
r_{\rm max} = \left. r_{\rm plateau}\text{\textcolor{white}{\Large j}}\right|_{T_H = T_{\rm max}}\,.
\ee
At this point, the PBH, in the last stage of its evaporation, is not able to provide enough energy to reheat the hot spot further.

\begin{figure}
\includegraphics[width=\linewidth]{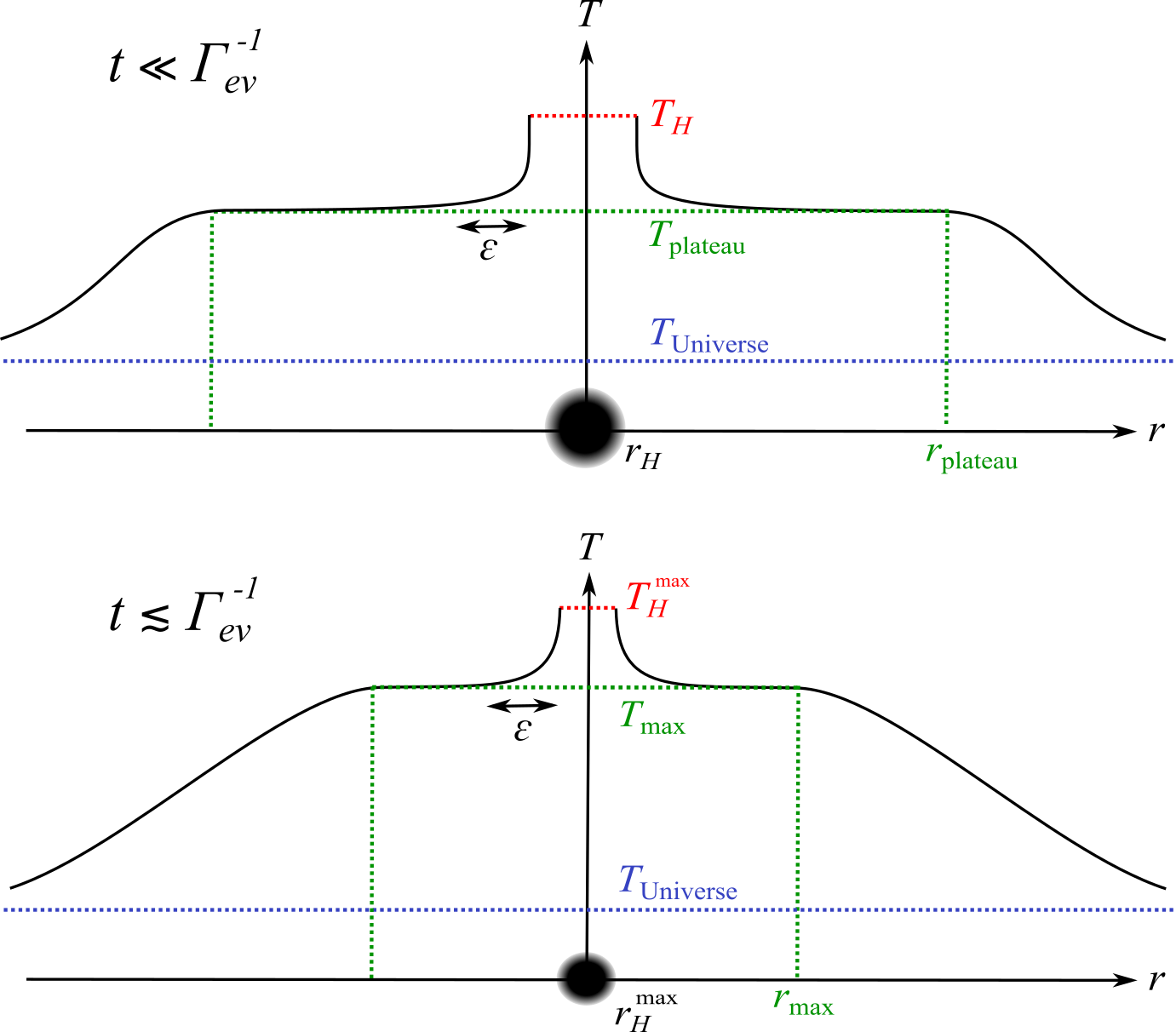}
\caption{\label{fig:profile}\footnotesize Sketch of the temperature profile around PBHs throughout the evaporation process.}
\end{figure}

To obtain these results, the authors of Ref.~\cite{He:2022wwy} assumed that Hawking radiation particles all deposited their energy at an averaged time $t_{\rm dep.}\equiv \Gamma(T)^{-1}$, where the first-splitting rate was estimated to be
 \be\label{eq:splitrate}
 \Gamma(T)\sim \alpha^2 T \sqrt{\frac{T}{T_H}}\,.
 \ee
We argue that this approximation cannot remain valid near the horizon. Indeed, even if the temperature within the hot spot was perfectly homogeneous, the energy deposition probability of Hawking radiation
\be
dP\sim \Gamma(T) e^{-(r-r_H)\Gamma(T)}dr\,,
\ee
is a decreasing function of the radius. In addition, the Hawking radiation energy density also decreases like $\propto r^{-2}$. Energy deposition is thus more efficient close to the horizon, leading to a local increase of the temperature. Because an increase of the local temperature also corresponds to an enhancement of the first-splitting rate of Eq.~\eqref{eq:splitrate}, the plasma temperature must interpolate between the Hawking temperature at the horizon and the temperature of the hot spot calculated on larger distances in Eq.~\eqref{eq:plateau} and \eqref{eq:max}. In Fig.~\ref{fig:profile}, we present the qualitative temperature profile that must be present around the hot spot at $t\ll\Gamma_{\rm ev}^{-1}$, when the hot spot core temperature was $T_{\rm plateau}$ (upper panel), and at $t\lesssim \Gamma_{\rm ev}^{-1}$, when the plateau temperature has saturated at $T_{\rm max}$ (lower panel).
Note that by enforcing the smooth behaviour of the plasma $T\rightarrow T_H$ as $t\rightarrow r_H$, we have introduced a physical realisation of the regularisation procedure of Ref.~\cite{Gregory:2013hja} that was necessary to evaluate the contribution of the conical deficit to the action in Eq.~\eqref{eq:conical}. In full generality, one would need to solve the heat equation in the vicinity of PBHs to obtain the exact temperature profile and, in particular, to evaluate the smoothing size $\varepsilon$ that is sent to zero to obtain Eq.~\ref{eq:RuthResult}. For simplicity, we will assume here that the typical size of this smoothing is negligible as compared to the typical size of a true vacuum bubble and leave a precise study of the profile geometry contribution for future work.

\bigskip
\paragraph{Bounce Solution and Nucleation Rate. ---}

Let us now consider the case of a homogeneous scalar-field configuration $\phi_+$ living in a metastable minimum of its potential where $V(\phi_+)=0$ before tunneling to the true vacuum, located at $\phi_-$, for which $V(\phi_-)<0$. To evaluate the nucleation rate of true-vacuum  bubbles, one needs to search for a saddle-point `bounce' solution of the action in Eq.~\eqref{eq:action}. In a finite temperature background, one also needs to account for thermal quantum corrections to the scalar potential, such that $V(\phi)=V(\phi, T)$. Following Ref.~\cite{Dai:2019eei}, we parametrise the metric as
\begin{equation}
ds^2=f(r)e^{2\delta(r)}d\tau^2+\frac{dr^2}{f(r)}+r^2 d\Omega_2^2\,,
\end{equation}
where $f(r)\equiv 1-2G\mu(r)/r$ with $\mu(r)$ representing the local black hole mass.
We then search for solutions of the classical equations of motion satisfying the boundary conditions\footnote{See Ref.~\cite{Dai:2019eei} for more details on the resolution.}
\begin{equation}
\lim_{r\rightarrow\infty}\phi(r)\rightarrow \phi_+\,\, ,\,\, M_+\equiv\lim_{r\rightarrow\infty}\mu(r)\,.
\end{equation}
Assuming that the energy of the bubble is entirely provided by the mass variation of the black hole inside the bubble, its mass at infinity is unchanged, and $M_+$ corresponds to the PBH mass before the transition. In the near horizon limit, due to the presence of a negative cosmological constant $\Lambda_- \equiv V(\phi_-)$, the mass parameter at the horizon $\mu_- \equiv \lim_{r\rightarrow r_H}\mu(r)$ slightly differs from the physical  (ADM) mass of the black hole inside the bubble, $M_-$. The Schwarzschild radius can be related to these masses by the relation
\be
r_H = 2G\mu_- = 2GM_- + \frac{\Lambda_- r_H^3}{3}\,. 
\ee
With such a solution in hand, one can calculate the bounce action of Eq.~\eqref{eq:action} numerically. Assuming a local thermal equilibrium around PBHs with a temperature profile $T(r)$ such that $\lim_{r\rightarrow r_H}T(r) = T_H$, the metric does not feature any conical singularity at the horizon, but the contribution of the effective conical deficit far from the horizon is entirely contained in its contribution to the Ricci scalar. For simplicity, we will assume that the interpolation distance $\varepsilon$ is small enough that the regularisation procedure described in \cite{Gregory:2013hja, Burda:2015yfa} is sufficient, and we will use equivalently the result of Eq.~\eqref{eq:RuthResult} to calculate the action. 

Despite the powerful generality of Eq.~\eqref{eq:RuthResult}, the calculation of the corresponding nucleation rate was always restricted to the case of the Hartle-Hawking vacuum~\cite{Burda:2015isa, Burda:2016mou, Burda:2015yfa, Gregory:2013hja, Dai:2019eei}. Using dimensional analysis to obtain the pre-factor, the nucleation rate was obtained in Ref.~\cite{Burda:2015isa, Burda:2016mou, Burda:2015yfa}  by simply writing 
\be
\Gamma_{\rm FVD}^{\rm HH}\equiv (GM_+)^{-1}\left(\frac{I_{\rm b}[T_H]}{2\pi}\right)^{1/2}\exp\left(-I_{\rm b}[T_H]\right)\,.
\ee
By noticing that $(GM_+)^{-1}\sim T_H$, one can interpret this result as the FVD rate in a thermal plasma derived in Ref.~\cite{Linde:1980tt}.  Generalizing this analogy to the case of a thermal hot spot with temperature $T<T_H$, we thus propose to calculate the nucleation rate using
\bea\label{eq:rate}
\Gamma_{\rm FVD}(T)&\approx& T \left(\frac{I_{\rm b}[T]}{2\pi }\right)^{1/2}\exp\left(-I_{\rm b}[T]\right)\,,\nonumber\\
&\approx & T \left(\frac{I_{\rm b}[T_H]}{2\pi }\right)^{1/2}\exp\left(-I_{\rm b}[T_H]\right)\,,
\eea
where we used Eq.~\eqref{eq:RuthResult} to go from the first line to the second. Note that in this equation, thanks to Eq.~\eqref{eq:RuthResult}, the only suppression is due to the fact that $T<T_H$ appears in the pre-factor, as the four-dimensional action does not depend on the hot-spot temperature value. 

Before we derive bounds on PBHs using this result, one needs to discuss what the actual value of the temperature $T$ should be when calculating the FVD rate in Eq.~\ref{eq:rate}. In the previous section, we have seen that PBHs with masses $M\gtrsim 0.8\mathrm{g}$ are surrounded for a long time by a hot spot with temperature $T_{\rm plateau}\approx 2\times 10^{-4}T_H$ before they evaporate and the hot spot reaches a maximum plateau temperature $T_{\rm max}\approx 2\times 10^9\mathrm{GeV}$ that is independent on the initial PBH mass. Once the hot spot has formed at $T_{\rm plateau}$, its temperature remains constant throughout most of the PBH lifetime, enabling the use of the Euclidean formalism in this regime. Therefore, one can safely use $T_{\rm plateau}$ to calculate the nucleation rate. 
However, such a choice is extremely conservative. Indeed, every PBH heavier than $M_\star$ is expected to evaporate and reach a point where $M=M_\star$ and the hot spot temperature reaches $T_{\rm max}$. Because $T_{\rm max}\geqslant T_{\rm plateau}$, 
\be
\left.I_b[T_H]\right|_{M = M_\star}\ll \left.I_b[T_H]\right|_{M > M_\star} 
\,,\ee
leading to a much larger FVD rate independent on the PBH mass. A legitimate concern is whether the Euclidean formalism is reliable given that the PBH and its environment is dynamical. To ensure this, we restrict ourselves to cases where the characteristic timescales are much larger than FVD, \textit{i.e.} $\Gamma_{\rm FVD}/\Gamma_{\rm ev}\gg 1$.

\paragraph{Case of the Electroweak Vacuum. --- \label{sec:bounds_PBH}}
Let us now apply our results to the case of the electroweak vacuum. In order to precisely calculate $\Gamma_{\rm FVD}$, we parameterise the Higgs potential, 
\begin{equation}
    V(\phi)=\lambda_\text{eff}(\phi)\phi^4/4 \,,
\end{equation}
where 
\begin{equation}
    \lambda_\text{eff}=\lambda_*+b\left(\ln\frac{\phi}{M_p}\right)^2+c\left(\ln\frac{\phi}{M_p}\right)^4\,,
\end{equation}
in which we determined the parameters $(\lambda_*,b,c)=(-3.2\times 10^{-3},\,-1.497\times 10^{-6},\,5.42\times 10^{-8})$ by taking the updated measurements of the top quark mass from the CMS collaboration~\cite{CMS:2023ebf} and using the numerical package PyR@te 3~\cite{Poole:2019kcm, Sartore:2020gou} to solve the renormalization group running at the three-loop level. To this zero-temperature potential, one needs to add one-loop thermal corrections:
\be
\Delta V(\phi,T) = \kappa^2 T_\text{plasma}^2\phi^2\,,
\ee
where $\kappa=0.35$ following (See Ref.~\cite{Strumia:2022jil} and ref therein). Given these parameters, the electroweak vacuum appears to be metastable, although its lifetime is much longer than the age of the Universe in the absence of evaporating PBHs~\cite{Degrassi:2014hoa,Alekhin:2012py,Espinosa:2015qea,Buttazzo:2013uya}. In Refs.~\cite{Gregory:2013hja, Burda:2015isa, Burda:2016mou, Hayashi:2020ocn, Burda:2015yfa}, the effect of PBHs on the FVD was used to derive constraints on the amount of PBHs formed in the early Universe, using the Hartle-Hawking vacuum, and ignoring thermal corrections to the potential. Here, we propose to use the result of Eq.~\eqref{eq:rate} to obtain more realistic results.
\begin{figure}
\includegraphics[width= \linewidth]{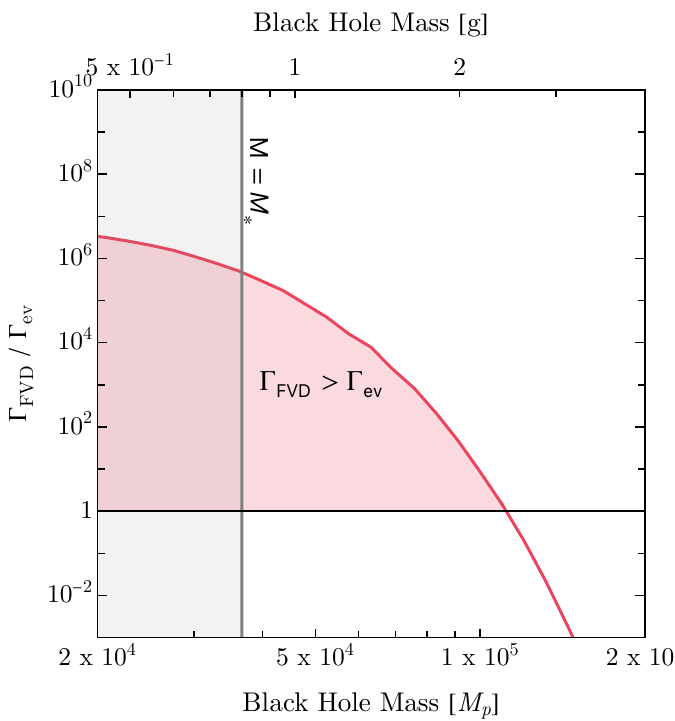}
\caption{\label{fig:ratio}\footnotesize Evolution of the ratio $\Gamma_{\rm FVD}/\Gamma_{\rm ev}$ for different values of the PBH mass, using the initial plateau temperature $T_{\rm plateau}$ to calculate $\Gamma_{\rm FVD}$ (red curve). The gray-shaded area depicts the region where $M<M_\star$, and the red-shaded area the region where the ratio is larger than unity.}
\end{figure}

\paragraph{PBH bounds and Discussion. ---} 
In Fig.~\ref{fig:ratio}, we depict the evolution of the ratio $\Gamma_{\rm FVD}/\Gamma_{\rm ev}$ as a function of the PBH mass. As one can see, 
$\Gamma_{\rm FVD}/\Gamma_{\rm ev}$ is larger than one for $M_\star < M \lesssim 2\mathrm{g}$. However, $\Gamma_{\rm FVD}$ is exponentially suppressed for larger PBH masses. Remarkably, the value of this ratio at $M=M_\star$ is of about $5\times 10^5\gg 1$. PBHs with large initial masses --- for which $\Gamma_{\rm FVD}(T_{\rm plateau})/\Gamma_{\rm ev}\ll 1$ --- thus feature a much larger bubble nucleation rate once they evaporate and reach the point where $M=M_\star$. The FVD is then a much faster process than the PBH evaporation at that point. For PBHs with initial masses in the region where $M<M_\star$, the hot spot does not have enough time to form in the first place~\cite{He:2022wwy}. However, PBHs with initial masses larger than $M_\star$, will eventually enter this region when they evaporate. After they pass the threshold $M=M_\star$, such PBHs do not have enough energy to reheat the hot spot. Therefore the hot spot temperature stays constant at $T=T_{\rm max}$ throughout the end of the evaporation. In this region, we therefore fixed the temperature at $T_{\rm max}$ in the calculation.

In Ref.~\cite{Strumia:2022jil}, the effects of thermal corrections were claimed to become sizeable for $M\lesssim 10^3\mathrm{g}$. However, such effects only become relevant for \mbox{$M\lesssim 0.1\mathrm{g}<M_\star$} in our case and do not affect any of the conclusions drawn in this {\em letter}.
Indeed, because we now consider the hot spot temperature to be $T_{\rm plateau}\approx 10^{-4}T_\text{H}$, the effect of thermal corrections is shifted to lower masses by a factor $\mathcal O(10^{-4})$, in agreement with our results. Another important feature of the bubble profiles we obtain is that they always have radii much smaller than $r_{\rm plateau}$ and $r_{\rm max}$. This validates the assumption of homogeneous plateau temperature that we have used throughout this work. We also checked that the value of the action calculated using Eq.~\eqref{eq:action} and substracting the background action and conical deficit matches exactly with the expression in Eq.~\eqref{eq:RuthResult}.


To convert the FVD rate around a black hole into a probability of FVD throughout cosmic history, one needs to evaluate time $\Delta t$ over which the FVD rate calculation remains valid during the PBH lifetime. The corresponding total decay probability can then be evaluated as
\be
P_{\rm FVD}\equiv 1-e^{-\Gamma_{\rm FVD} \Delta t}\,.
\ee
First, we have seen that the hot spot temperature stays constant at $T=T_{\rm plateau}$ during most of the PBH lifetime $t_{\rm ev}$. Therefore, the false vacuum has about $\Delta t \sim \Gamma_{\rm ev}^{-1}$ to tunnel. In that case, 
\be
P_{\rm FVD}(M) = 1 - e^{-\Gamma_{\rm  FVD}(T_{\rm plateau})/\Gamma_{\rm ev}}\,.
\ee
Depending on the ratio $\Gamma_{\rm FVD}/\Gamma_{\rm ev}$ given in Fig.~\ref{fig:ratio}, this probability can vary over orders of magnitude. Once the black hole starts evaporating and eventually reaches $M=M_\star$, then we have seen that $\Gamma_{\rm FVD}/\Gamma_{\rm ev}\approx 5\times 10^5$. This ensures that $P_{\rm FVD}\approx 1$ as long as one considers time scales $\Delta t\lesssim 10^{-6} \times \Gamma_{\rm ev}^{-1}$. Over such a short time, the PBH mass variation is negligible, and the Euclidean formalism still holds.  
Finally, with these decay probabilities, we may now look at the ensuing bounds on the share of the earlier universe energy which collapsed into PBHs in cosmic history $\beta_{\rm PBH}=\rho_\text{PBH}/\rho_{\rm tot}$.  At the earliest, PBHs formed when an overdensity of size comparable to that of a Hubble patch collapsed gravitationally. Assuming the Universe to be radiation-dominated during that time such that $\rho_{\rm tot} = \rho_{\rm rad}$ one can calculate the Universe's temperature at formation to be
\be
T_f=\left(\frac{\gamma}{4\pi}\sqrt{\frac{45}{\pi g_\star(T_f)}}\frac{M_p^3}{M}\right)^{1/2}\,,
\ee
and obtain the value of the density fraction at formation
\be
\beta_{\rm PBH} =\frac{4}{3}\frac{MN_{\rm PBH}H_0^3}{s_0 T_f} \approx 2 \times 10^{-80}N_{\rm PBH} \left(\frac{M}{M_\star}\right)^{3/2}\,,
\ee
where $\gamma=(1/\sqrt{3})^3$ is a numerical factor related to the gravitational collapse~\cite{Carr:1975qj, Carr:2009jm}, $N_{\rm PBH}$ is the total number of PBHs that, if stable, would be contained in a Hubble patch of size $H_0^{-1}$, where $s_0 \approx 2\times 10^{-38}\mathrm{GeV^3}$ is the entropy density of the Universe today, and $H_0 \approx 70\ \mathrm{km^{-1}s^{-1}Mpc^{-1}}$ is the Hubble constant today. 

\begin{figure}
\includegraphics[width = \linewidth]{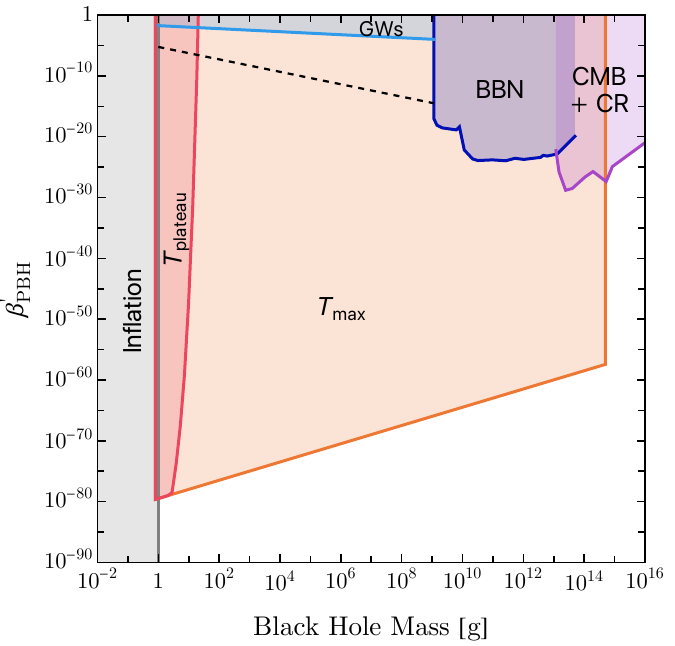}
\caption{\label{fig:beta}\footnotesize Constraints on the PBH energy fraction at formation, \mbox{$\beta_{\rm PBH}'\equiv \gamma^{1/2}(g_\star(T_f)/106.75)^{-1/4}\beta_{\rm PBH}$},  as a function of its mass. The orange-shaded region represents the most competitive constraint derived using $T_{\rm max}$ for the hot spot temperature, whereas the red-shaded region denotes the most conservative bound using $T_{\rm plateau}$.}
\end{figure}
Using a Poisson distribution, demanding that the electroweak vacuum has never decayed in a Hubble patch until today at $95\%$ confidence level is equivalent to requesting that 
\begin{equation}
    N_\text{PBH}\,P_d < 2.7\,,
\end{equation}
leading, in the case where $P_d\approx 1$, to the constraint
\be
\beta_{\rm PBH} \lesssim 5 \times 10^{-80} \left(\frac{M}{M_\star}\right)^{3/2}\,.
\ee
In Fig.~\ref{fig:beta}, we draw the corresponding constraints both when considering $T_{\rm plateau}$ and $T_{\rm max}$ for the hot spot temperature.  Note that for masses larger than $M_{\star}$ but smaller than $88$g ($4$g), the temperature $T_{\rm plateau}$ ($T_{\rm max}$) is smaller than the plasma temperature at the time of evaporation. In this case, we slightly modified the result of Eq.~\eqref{eq:rate}, by  using the plasma temperature at evaporation instead of the hot spot temperature:
\be
\Gamma_{\rm FVD}(T)\approx \max\{T, T_{\rm ev}\} \left(\frac{I_{\rm b}[T_H]}{2\pi }\right)^{1/2}\exp\left(-I_{\rm b}[T_H]\right)\,,
\ee
where
\be
T_{\rm ev}\equiv \left(\frac{90}{8\pi^3 g_\star(t_{\rm ev})}\right)^{1/4}\!\!\sqrt{\Gamma_{\rm ev}M_p}\,,
\ee
is the plasma temperature at evaporation. In the regime where such constraints are relevant, we checked that the ratio $\Gamma_{\rm FVD}/\Gamma_{\rm ev}\gg 1$ guaranteeing the validity of the Euclidean formalism. Note also that thermal corrections in the action, arising now at $T_{\rm ev}$ rather than $T_{\rm plateau}$ or $T_{\rm max}$, remain subdominant in this region of parameter space.
For comparison, we indicate other constraints from inflation, Big Bang Nucleosynthesis, CMB distortion and $\gamma$-rays~\cite{Carr:2020gox}.


The conclusion one can draw from the results above is twofold: If the electroweak vacuum is confirmed to be metastable with future SM measurements, such limits exclude many scenarios predicting a large abundance of evaporating PBHs in cosmology~\cite{Inomata:2020lmk,Barman:2022pdo,Bhaumik:2022zdd,Morrison:2018xla,Gondolo:2020uqv,Bernal:2020ili,JyotiDas:2021shi,Hamada:2016jnq,Hooper:2020otu,Chaudhuri:2020wjo,Kitabayashi:2021hox,Cheek:2021odj,Martin:2019nuw,Martin:2020fgl,Auclair:2020csm,Cheek:2022dbx,Hooper:2019gtx,Cheek:2021cfe,Perez-Gonzalez:2020vnz,Bernal:2020kse,Datta:2020bht,Masina:2021zpu,Bernal:2020bjf,Baldes:2020nuv,Allahverdi:2017sks,Masina:2020xhk,Lunardini:2019zob,Dienes:2022zgd,Papanikolaou:2022chm,Saito:2008jc}. In particular, they would exclude entirely the possibility that PBHs dominate the Universe before evaporating (above the dotted line in Fig.~\ref{fig:beta}). Alternatively, if such scenarios are confirmed using cosmological data, our results may be used as a strong indication that new physics is required to stabilise the electroweak vacuum throughout cosmic history.
Finally, we emphasize that the results presented in this {\em letter} can be applied to any first-order phase transition taking place in cosmology.

%
%



\bigskip
\bigskip


\paragraph{Acknowledgments. ---}

The authors would like to thank W.-Y.~Ai for extremely useful discussions throughout the realisation of this work, as well as R.~Gregory for providing us with helpful feedback on the manuscript.
L.~Hamaide is supported by the King's Cromwell Scholarship. S.~Hu is supported by the Chinese Scholarship Council and King's College London. The work of L.~Heurtier is supported in part by the U.K.\ Science and Technology Facilities Council (STFC) 
under Grant ST/P001246/1. A.~Cheek is supported by the grant ``AstroCeNT: Particle Astrophysics Science and Technology Centre" carried out within the International Research Agendas program of the Foundation for Polish Science financed by the European Union under the European Regional Development Fund. L. Heurtier would like to thank his recently born daughter, Lilie Ming-Xuan Heurtier, for providing him with the extreme happiness required to writing this letter despite the lack of sleep.


\bigskip

\bibliography{draft.bib}

\begin{thebibliography}{59}%
\makeatletter
\providecommand \@ifxundefined [1]{%
 \@ifx{#1\undefined}
}%
\providecommand \@ifnum [1]{%
 \ifnum #1\expandafter \@firstoftwo
 \else \expandafter \@secondoftwo
 \fi
}%
\providecommand \@ifx [1]{%
 \ifx #1\expandafter \@firstoftwo
 \else \expandafter \@secondoftwo
 \fi
}%
\providecommand \natexlab [1]{#1}%
\providecommand \enquote  [1]{``#1''}%
\providecommand \bibnamefont  [1]{#1}%
\providecommand \bibfnamefont [1]{#1}%
\providecommand \citenamefont [1]{#1}%
\providecommand \href@noop [0]{\@secondoftwo}%
\providecommand \href [0]{\begingroup \@sanitize@url \@href}%
\providecommand \@href[1]{\@@startlink{#1}\@@href}%
\providecommand \@@href[1]{\endgroup#1\@@endlink}%
\providecommand \@sanitize@url [0]{\catcode `\\12\catcode `\$12\catcode
  `\&12\catcode `\#12\catcode `\^12\catcode `\_12\catcode `\%12\relax}%
\providecommand \@@startlink[1]{}%
\providecommand \@@endlink[0]{}%
\providecommand \url  [0]{\begingroup\@sanitize@url \@url }%
\providecommand \@url [1]{\endgroup\@href {#1}{\urlprefix }}%
\providecommand \urlprefix  [0]{URL }%
\providecommand \Eprint [0]{\href }%
\providecommand \doibase [0]{http://dx.doi.org/}%
\providecommand \selectlanguage [0]{\@gobble}%
\providecommand \bibinfo  [0]{\@secondoftwo}%
\providecommand \bibfield  [0]{\@secondoftwo}%
\providecommand \translation [1]{[#1]}%
\providecommand \BibitemOpen [0]{}%
\providecommand \bibitemStop [0]{}%
\providecommand \bibitemNoStop [0]{.\EOS\space}%
\providecommand \EOS [0]{\spacefactor3000\relax}%
\providecommand \BibitemShut  [1]{\csname bibitem#1\endcsname}%
\let\auto@bib@innerbib\@empty
\bibitem [{\citenamefont {Hartle}\ and\ \citenamefont
  {Hawking}(1976)}]{Hartle:1976tp}%
  \BibitemOpen
  \bibfield  {author} {\bibinfo {author} {\bibfnamefont {J.~B.}\ \bibnamefont
  {Hartle}}\ and\ \bibinfo {author} {\bibfnamefont {S.~W.}\ \bibnamefont
  {Hawking}},\ }\href {\doibase 10.1103/PhysRevD.13.2188} {\bibfield  {journal}
  {\bibinfo  {journal} {Phys. Rev. D}\ }\textbf {\bibinfo {volume} {13}},\
  \bibinfo {pages} {2188} (\bibinfo {year} {1976})}\BibitemShut {NoStop}%
\bibitem [{\citenamefont {Gregory}\ \emph {et~al.}(2014)\citenamefont
  {Gregory}, \citenamefont {Moss},\ and\ \citenamefont
  {Withers}}]{Gregory:2013hja}%
  \BibitemOpen
  \bibfield  {author} {\bibinfo {author} {\bibfnamefont {R.}~\bibnamefont
  {Gregory}}, \bibinfo {author} {\bibfnamefont {I.~G.}\ \bibnamefont {Moss}}, \
  and\ \bibinfo {author} {\bibfnamefont {B.}~\bibnamefont {Withers}},\ }\href
  {\doibase 10.1007/JHEP03(2014)081} {\bibfield  {journal} {\bibinfo  {journal}
  {JHEP}\ }\textbf {\bibinfo {volume} {03}},\ \bibinfo {pages} {081} (\bibinfo
  {year} {2014})},\ \Eprint {http://arxiv.org/abs/1401.0017} {arXiv:1401.0017
  [hep-th]} \BibitemShut {NoStop}%
\bibitem [{\citenamefont {Burda}\ \emph
  {et~al.}(2015{\natexlab{a}})\citenamefont {Burda}, \citenamefont {Gregory},\
  and\ \citenamefont {Moss}}]{Burda:2015isa}%
  \BibitemOpen
  \bibfield  {author} {\bibinfo {author} {\bibfnamefont {P.}~\bibnamefont
  {Burda}}, \bibinfo {author} {\bibfnamefont {R.}~\bibnamefont {Gregory}}, \
  and\ \bibinfo {author} {\bibfnamefont {I.}~\bibnamefont {Moss}},\ }\href
  {\doibase 10.1103/PhysRevLett.115.071303} {\bibfield  {journal} {\bibinfo
  {journal} {Phys. Rev. Lett.}\ }\textbf {\bibinfo {volume} {115}},\ \bibinfo
  {pages} {071303} (\bibinfo {year} {2015}{\natexlab{a}})},\ \Eprint
  {http://arxiv.org/abs/1501.04937} {arXiv:1501.04937 [hep-th]} \BibitemShut
  {NoStop}%
\bibitem [{\citenamefont {Burda}\ \emph
  {et~al.}(2015{\natexlab{b}})\citenamefont {Burda}, \citenamefont {Gregory},\
  and\ \citenamefont {Moss}}]{Burda:2015yfa}%
  \BibitemOpen
  \bibfield  {author} {\bibinfo {author} {\bibfnamefont {P.}~\bibnamefont
  {Burda}}, \bibinfo {author} {\bibfnamefont {R.}~\bibnamefont {Gregory}}, \
  and\ \bibinfo {author} {\bibfnamefont {I.}~\bibnamefont {Moss}},\ }\href
  {\doibase 10.1007/JHEP08(2015)114} {\bibfield  {journal} {\bibinfo  {journal}
  {JHEP}\ }\textbf {\bibinfo {volume} {08}},\ \bibinfo {pages} {114} (\bibinfo
  {year} {2015}{\natexlab{b}})},\ \Eprint {http://arxiv.org/abs/1503.07331}
  {arXiv:1503.07331 [hep-th]} \BibitemShut {NoStop}%
\bibitem [{\citenamefont {Burda}\ \emph {et~al.}(2016)\citenamefont {Burda},
  \citenamefont {Gregory},\ and\ \citenamefont {Moss}}]{Burda:2016mou}%
  \BibitemOpen
  \bibfield  {author} {\bibinfo {author} {\bibfnamefont {P.}~\bibnamefont
  {Burda}}, \bibinfo {author} {\bibfnamefont {R.}~\bibnamefont {Gregory}}, \
  and\ \bibinfo {author} {\bibfnamefont {I.}~\bibnamefont {Moss}},\ }\href
  {\doibase 10.1007/JHEP06(2016)025} {\bibfield  {journal} {\bibinfo  {journal}
  {JHEP}\ }\textbf {\bibinfo {volume} {06}},\ \bibinfo {pages} {025} (\bibinfo
  {year} {2016})},\ \Eprint {http://arxiv.org/abs/1601.02152} {arXiv:1601.02152
  [hep-th]} \BibitemShut {NoStop}%
\bibitem [{\citenamefont {Hiller}\ \emph {et~al.}(2022)\citenamefont {Hiller},
  \citenamefont {H\"ohne}, \citenamefont {Litim},\ and\ \citenamefont
  {Steudtner}}]{Hiller:2022rla}%
  \BibitemOpen
  \bibfield  {author} {\bibinfo {author} {\bibfnamefont {G.}~\bibnamefont
  {Hiller}}, \bibinfo {author} {\bibfnamefont {T.}~\bibnamefont {H\"ohne}},
  \bibinfo {author} {\bibfnamefont {D.~F.}\ \bibnamefont {Litim}}, \ and\
  \bibinfo {author} {\bibfnamefont {T.}~\bibnamefont {Steudtner}},\ }\href
  {\doibase 10.1103/PhysRevD.106.115004} {\bibfield  {journal} {\bibinfo
  {journal} {Phys. Rev. D}\ }\textbf {\bibinfo {volume} {106}},\ \bibinfo
  {pages} {115004} (\bibinfo {year} {2022})},\ \Eprint
  {http://arxiv.org/abs/2207.07737} {arXiv:2207.07737 [hep-ph]} \BibitemShut
  {NoStop}%
\bibitem [{\citenamefont {Unruh}(1976)}]{Unruh:1976db}%
  \BibitemOpen
  \bibfield  {author} {\bibinfo {author} {\bibfnamefont {W.~G.}\ \bibnamefont
  {Unruh}},\ }\href {\doibase 10.1103/PhysRevD.14.870} {\bibfield  {journal}
  {\bibinfo  {journal} {Phys. Rev. D}\ }\textbf {\bibinfo {volume} {14}},\
  \bibinfo {pages} {870} (\bibinfo {year} {1976})}\BibitemShut {NoStop}%
\bibitem [{\citenamefont {Kohri}\ and\ \citenamefont
  {Matsui}(2018)}]{Kohri:2017ybt}%
  \BibitemOpen
  \bibfield  {author} {\bibinfo {author} {\bibfnamefont {K.}~\bibnamefont
  {Kohri}}\ and\ \bibinfo {author} {\bibfnamefont {H.}~\bibnamefont {Matsui}},\
  }\href {\doibase 10.1103/PhysRevD.98.123509} {\bibfield  {journal} {\bibinfo
  {journal} {Phys. Rev. D}\ }\textbf {\bibinfo {volume} {98}},\ \bibinfo
  {pages} {123509} (\bibinfo {year} {2018})},\ \Eprint
  {http://arxiv.org/abs/1708.02138} {arXiv:1708.02138 [hep-ph]} \BibitemShut
  {NoStop}%
\bibitem [{\citenamefont {Hayashi}\ \emph {et~al.}(2020)\citenamefont
  {Hayashi}, \citenamefont {Kamada}, \citenamefont {Oshita},\ and\
  \citenamefont {Yokoyama}}]{Hayashi:2020ocn}%
  \BibitemOpen
  \bibfield  {author} {\bibinfo {author} {\bibfnamefont {T.}~\bibnamefont
  {Hayashi}}, \bibinfo {author} {\bibfnamefont {K.}~\bibnamefont {Kamada}},
  \bibinfo {author} {\bibfnamefont {N.}~\bibnamefont {Oshita}}, \ and\ \bibinfo
  {author} {\bibfnamefont {J.}~\bibnamefont {Yokoyama}},\ }\href {\doibase
  10.1007/JHEP08(2020)088} {\bibfield  {journal} {\bibinfo  {journal} {JHEP}\
  }\textbf {\bibinfo {volume} {08}},\ \bibinfo {pages} {088} (\bibinfo {year}
  {2020})},\ \Eprint {http://arxiv.org/abs/2005.12808} {arXiv:2005.12808
  [hep-th]} \BibitemShut {NoStop}%
\bibitem [{\citenamefont {Shkerin}\ and\ \citenamefont
  {Sibiryakov}(2021)}]{Shkerin:2021zbf}%
  \BibitemOpen
  \bibfield  {author} {\bibinfo {author} {\bibfnamefont {A.}~\bibnamefont
  {Shkerin}}\ and\ \bibinfo {author} {\bibfnamefont {S.}~\bibnamefont
  {Sibiryakov}},\ }\href {\doibase 10.1007/JHEP11(2021)197} {\bibfield
  {journal} {\bibinfo  {journal} {JHEP}\ }\textbf {\bibinfo {volume} {11}},\
  \bibinfo {pages} {197} (\bibinfo {year} {2021})},\ \Eprint
  {http://arxiv.org/abs/2105.09331} {arXiv:2105.09331 [hep-th]} \BibitemShut
  {NoStop}%
\bibitem [{\citenamefont {Shkerin}\ and\ \citenamefont
  {Sibiryakov}(2022)}]{Shkerin:2021rhy}%
  \BibitemOpen
  \bibfield  {author} {\bibinfo {author} {\bibfnamefont {A.}~\bibnamefont
  {Shkerin}}\ and\ \bibinfo {author} {\bibfnamefont {S.}~\bibnamefont
  {Sibiryakov}},\ }\href {\doibase 10.1007/JHEP08(2022)161} {\bibfield
  {journal} {\bibinfo  {journal} {JHEP}\ }\textbf {\bibinfo {volume} {08}},\
  \bibinfo {pages} {161} (\bibinfo {year} {2022})},\ \Eprint
  {http://arxiv.org/abs/2111.08017} {arXiv:2111.08017 [hep-th]} \BibitemShut
  {NoStop}%
\bibitem [{\citenamefont {Strumia}(2022)}]{Strumia:2022jil}%
  \BibitemOpen
  \bibfield  {author} {\bibinfo {author} {\bibfnamefont {A.}~\bibnamefont
  {Strumia}},\ }\href@noop {} {\  (\bibinfo {year} {2022})},\ \Eprint
  {http://arxiv.org/abs/2209.05504} {arXiv:2209.05504 [hep-ph]} \BibitemShut
  {NoStop}%
\bibitem [{\citenamefont {Briaud}\ \emph {et~al.}(2022)\citenamefont {Briaud},
  \citenamefont {Shkerin},\ and\ \citenamefont {Sibiryakov}}]{Briaud:2022few}%
  \BibitemOpen
  \bibfield  {author} {\bibinfo {author} {\bibfnamefont {V.}~\bibnamefont
  {Briaud}}, \bibinfo {author} {\bibfnamefont {A.}~\bibnamefont {Shkerin}}, \
  and\ \bibinfo {author} {\bibfnamefont {S.}~\bibnamefont {Sibiryakov}},\
  }\href {\doibase 10.1103/PhysRevD.106.125001} {\bibfield  {journal} {\bibinfo
   {journal} {Phys. Rev. D}\ }\textbf {\bibinfo {volume} {106}},\ \bibinfo
  {pages} {125001} (\bibinfo {year} {2022})},\ \Eprint
  {http://arxiv.org/abs/2210.08028} {arXiv:2210.08028 [gr-qc]} \BibitemShut
  {NoStop}%
\bibitem [{\citenamefont {Das}\ and\ \citenamefont {Hook}(2021)}]{Das:2021wei}%
  \BibitemOpen
  \bibfield  {author} {\bibinfo {author} {\bibfnamefont {S.}~\bibnamefont
  {Das}}\ and\ \bibinfo {author} {\bibfnamefont {A.}~\bibnamefont {Hook}},\
  }\href {\doibase 10.1007/JHEP12(2021)145} {\bibfield  {journal} {\bibinfo
  {journal} {JHEP}\ }\textbf {\bibinfo {volume} {12}},\ \bibinfo {pages} {145}
  (\bibinfo {year} {2021})},\ \Eprint {http://arxiv.org/abs/2109.00039}
  {arXiv:2109.00039 [hep-ph]} \BibitemShut {NoStop}%
\bibitem [{\citenamefont {He}\ \emph {et~al.}(2023)\citenamefont {He},
  \citenamefont {Kohri}, \citenamefont {Mukaida},\ and\ \citenamefont
  {Yamada}}]{He:2022wwy}%
  \BibitemOpen
  \bibfield  {author} {\bibinfo {author} {\bibfnamefont {M.}~\bibnamefont
  {He}}, \bibinfo {author} {\bibfnamefont {K.}~\bibnamefont {Kohri}}, \bibinfo
  {author} {\bibfnamefont {K.}~\bibnamefont {Mukaida}}, \ and\ \bibinfo
  {author} {\bibfnamefont {M.}~\bibnamefont {Yamada}},\ }\href {\doibase
  10.1088/1475-7516/2023/01/027} {\bibfield  {journal} {\bibinfo  {journal}
  {JCAP}\ }\textbf {\bibinfo {volume} {01}},\ \bibinfo {pages} {027} (\bibinfo
  {year} {2023})},\ \Eprint {http://arxiv.org/abs/2210.06238} {arXiv:2210.06238
  [hep-ph]} \BibitemShut {NoStop}%
\bibitem [{\citenamefont {Dai}\ \emph {et~al.}(2020)\citenamefont {Dai},
  \citenamefont {Gregory},\ and\ \citenamefont {Stojkovic}}]{Dai:2019eei}%
  \BibitemOpen
  \bibfield  {author} {\bibinfo {author} {\bibfnamefont {D.-C.}\ \bibnamefont
  {Dai}}, \bibinfo {author} {\bibfnamefont {R.}~\bibnamefont {Gregory}}, \ and\
  \bibinfo {author} {\bibfnamefont {D.}~\bibnamefont {Stojkovic}},\ }\href
  {\doibase 10.1103/PhysRevD.101.125012} {\bibfield  {journal} {\bibinfo
  {journal} {Phys. Rev. D}\ }\textbf {\bibinfo {volume} {101}},\ \bibinfo
  {pages} {125012} (\bibinfo {year} {2020})},\ \Eprint
  {http://arxiv.org/abs/1909.00773} {arXiv:1909.00773 [hep-ph]} \BibitemShut
  {NoStop}%
\bibitem [{\citenamefont {MacGibbon}\ and\ \citenamefont
  {Webber}(1990)}]{MacGibbon:1990zk}%
  \BibitemOpen
  \bibfield  {author} {\bibinfo {author} {\bibfnamefont {J.~H.}\ \bibnamefont
  {MacGibbon}}\ and\ \bibinfo {author} {\bibfnamefont {B.~R.}\ \bibnamefont
  {Webber}},\ }\href {\doibase 10.1103/PhysRevD.41.3052} {\bibfield  {journal}
  {\bibinfo  {journal} {Phys. Rev. D}\ }\textbf {\bibinfo {volume} {41}},\
  \bibinfo {pages} {3052} (\bibinfo {year} {1990})}\BibitemShut {NoStop}%
\bibitem [{\citenamefont {MacGibbon}(1991)}]{MacGibbon:1991tj}%
  \BibitemOpen
  \bibfield  {author} {\bibinfo {author} {\bibfnamefont {J.~H.}\ \bibnamefont
  {MacGibbon}},\ }\href {\doibase 10.1103/PhysRevD.44.376} {\bibfield
  {journal} {\bibinfo  {journal} {Phys. Rev. D}\ }\textbf {\bibinfo {volume}
  {44}},\ \bibinfo {pages} {376} (\bibinfo {year} {1991})}\BibitemShut
  {NoStop}%
\bibitem [{\citenamefont {Linde}(1981)}]{Linde:1980tt}%
  \BibitemOpen
  \bibfield  {author} {\bibinfo {author} {\bibfnamefont {A.~D.}\ \bibnamefont
  {Linde}},\ }\href {\doibase 10.1016/0370-2693(81)90281-1} {\bibfield
  {journal} {\bibinfo  {journal} {Phys. Lett. B}\ }\textbf {\bibinfo {volume}
  {100}},\ \bibinfo {pages} {37} (\bibinfo {year} {1981})}\BibitemShut
  {NoStop}%
\bibitem [{CMS(2023)}]{CMS:2023ebf}%
  \BibitemOpen
  \href@noop {} {\  (\bibinfo {year} {2023})},\ \Eprint
  {http://arxiv.org/abs/2302.01967} {arXiv:2302.01967 [hep-ex]} \BibitemShut
  {NoStop}%
\bibitem [{\citenamefont {Poole}\ and\ \citenamefont
  {Thomsen}(2019)}]{Poole:2019kcm}%
  \BibitemOpen
  \bibfield  {author} {\bibinfo {author} {\bibfnamefont {C.}~\bibnamefont
  {Poole}}\ and\ \bibinfo {author} {\bibfnamefont {A.~E.}\ \bibnamefont
  {Thomsen}},\ }\href {\doibase 10.1007/JHEP09(2019)055} {\bibfield  {journal}
  {\bibinfo  {journal} {JHEP}\ }\textbf {\bibinfo {volume} {09}},\ \bibinfo
  {pages} {055} (\bibinfo {year} {2019})},\ \Eprint
  {http://arxiv.org/abs/1906.04625} {arXiv:1906.04625 [hep-th]} \BibitemShut
  {NoStop}%
\bibitem [{\citenamefont {Sartore}\ and\ \citenamefont
  {Schienbein}(2021)}]{Sartore:2020gou}%
  \BibitemOpen
  \bibfield  {author} {\bibinfo {author} {\bibfnamefont {L.}~\bibnamefont
  {Sartore}}\ and\ \bibinfo {author} {\bibfnamefont {I.}~\bibnamefont
  {Schienbein}},\ }\href {\doibase 10.1016/j.cpc.2020.107819} {\bibfield
  {journal} {\bibinfo  {journal} {Comput. Phys. Commun.}\ }\textbf {\bibinfo
  {volume} {261}},\ \bibinfo {pages} {107819} (\bibinfo {year} {2021})},\
  \Eprint {http://arxiv.org/abs/2007.12700} {arXiv:2007.12700 [hep-ph]}
  \BibitemShut {NoStop}%
\bibitem [{\citenamefont {Degrassi}(2014)}]{Degrassi:2014hoa}%
  \BibitemOpen
  \bibfield  {author} {\bibinfo {author} {\bibfnamefont {G.}~\bibnamefont
  {Degrassi}},\ }\href {\doibase 10.1393/ncc/i2014-11735-1} {\bibfield
  {journal} {\bibinfo  {journal} {Nuovo Cim. C}\ }\textbf {\bibinfo {volume}
  {037}},\ \bibinfo {pages} {47} (\bibinfo {year} {2014})},\ \Eprint
  {http://arxiv.org/abs/1405.6852} {arXiv:1405.6852 [hep-ph]} \BibitemShut
  {NoStop}%
\bibitem [{\citenamefont {Alekhin}\ \emph {et~al.}(2012)\citenamefont
  {Alekhin}, \citenamefont {Djouadi},\ and\ \citenamefont
  {Moch}}]{Alekhin:2012py}%
  \BibitemOpen
  \bibfield  {author} {\bibinfo {author} {\bibfnamefont {S.}~\bibnamefont
  {Alekhin}}, \bibinfo {author} {\bibfnamefont {A.}~\bibnamefont {Djouadi}}, \
  and\ \bibinfo {author} {\bibfnamefont {S.}~\bibnamefont {Moch}},\ }\href
  {\doibase 10.1016/j.physletb.2012.08.024} {\bibfield  {journal} {\bibinfo
  {journal} {Phys. Lett. B}\ }\textbf {\bibinfo {volume} {716}},\ \bibinfo
  {pages} {214} (\bibinfo {year} {2012})},\ \Eprint
  {http://arxiv.org/abs/1207.0980} {arXiv:1207.0980 [hep-ph]} \BibitemShut
  {NoStop}%
\bibitem [{\citenamefont {Espinosa}\ \emph {et~al.}(2015)\citenamefont
  {Espinosa}, \citenamefont {Giudice}, \citenamefont {Morgante}, \citenamefont
  {Riotto}, \citenamefont {Senatore}, \citenamefont {Strumia},\ and\
  \citenamefont {Tetradis}}]{Espinosa:2015qea}%
  \BibitemOpen
  \bibfield  {author} {\bibinfo {author} {\bibfnamefont {J.~R.}\ \bibnamefont
  {Espinosa}}, \bibinfo {author} {\bibfnamefont {G.~F.}\ \bibnamefont
  {Giudice}}, \bibinfo {author} {\bibfnamefont {E.}~\bibnamefont {Morgante}},
  \bibinfo {author} {\bibfnamefont {A.}~\bibnamefont {Riotto}}, \bibinfo
  {author} {\bibfnamefont {L.}~\bibnamefont {Senatore}}, \bibinfo {author}
  {\bibfnamefont {A.}~\bibnamefont {Strumia}}, \ and\ \bibinfo {author}
  {\bibfnamefont {N.}~\bibnamefont {Tetradis}},\ }\href {\doibase
  10.1007/JHEP09(2015)174} {\bibfield  {journal} {\bibinfo  {journal} {JHEP}\
  }\textbf {\bibinfo {volume} {09}},\ \bibinfo {pages} {174} (\bibinfo {year}
  {2015})},\ \Eprint {http://arxiv.org/abs/1505.04825} {arXiv:1505.04825
  [hep-ph]} \BibitemShut {NoStop}%
\bibitem [{\citenamefont {Buttazzo}\ \emph {et~al.}(2013)\citenamefont
  {Buttazzo}, \citenamefont {Degrassi}, \citenamefont {Giardino}, \citenamefont
  {Giudice}, \citenamefont {Sala}, \citenamefont {Salvio},\ and\ \citenamefont
  {Strumia}}]{Buttazzo:2013uya}%
  \BibitemOpen
  \bibfield  {author} {\bibinfo {author} {\bibfnamefont {D.}~\bibnamefont
  {Buttazzo}}, \bibinfo {author} {\bibfnamefont {G.}~\bibnamefont {Degrassi}},
  \bibinfo {author} {\bibfnamefont {P.~P.}\ \bibnamefont {Giardino}}, \bibinfo
  {author} {\bibfnamefont {G.~F.}\ \bibnamefont {Giudice}}, \bibinfo {author}
  {\bibfnamefont {F.}~\bibnamefont {Sala}}, \bibinfo {author} {\bibfnamefont
  {A.}~\bibnamefont {Salvio}}, \ and\ \bibinfo {author} {\bibfnamefont
  {A.}~\bibnamefont {Strumia}},\ }\href {\doibase 10.1007/JHEP12(2013)089}
  {\bibfield  {journal} {\bibinfo  {journal} {JHEP}\ }\textbf {\bibinfo
  {volume} {12}},\ \bibinfo {pages} {089} (\bibinfo {year} {2013})},\ \Eprint
  {http://arxiv.org/abs/1307.3536} {arXiv:1307.3536 [hep-ph]} \BibitemShut
  {NoStop}%
\bibitem [{\citenamefont {Carr}(1975)}]{Carr:1975qj}%
  \BibitemOpen
  \bibfield  {author} {\bibinfo {author} {\bibfnamefont {B.~J.}\ \bibnamefont
  {Carr}},\ }\href {\doibase 10.1086/153853} {\bibfield  {journal} {\bibinfo
  {journal} {Astrophys. J.}\ }\textbf {\bibinfo {volume} {201}},\ \bibinfo
  {pages} {1} (\bibinfo {year} {1975})}\BibitemShut {NoStop}%
\bibitem [{\citenamefont {Carr}\ \emph {et~al.}(2010)\citenamefont {Carr},
  \citenamefont {Kohri}, \citenamefont {Sendouda},\ and\ \citenamefont
  {Yokoyama}}]{Carr:2009jm}%
  \BibitemOpen
  \bibfield  {author} {\bibinfo {author} {\bibfnamefont {B.~J.}\ \bibnamefont
  {Carr}}, \bibinfo {author} {\bibfnamefont {K.}~\bibnamefont {Kohri}},
  \bibinfo {author} {\bibfnamefont {Y.}~\bibnamefont {Sendouda}}, \ and\
  \bibinfo {author} {\bibfnamefont {J.}~\bibnamefont {Yokoyama}},\ }\href
  {\doibase 10.1103/PhysRevD.81.104019} {\bibfield  {journal} {\bibinfo
  {journal} {Phys. Rev. D}\ }\textbf {\bibinfo {volume} {81}},\ \bibinfo
  {pages} {104019} (\bibinfo {year} {2010})},\ \Eprint
  {http://arxiv.org/abs/0912.5297} {arXiv:0912.5297 [astro-ph.CO]} \BibitemShut
  {NoStop}%
\bibitem [{\citenamefont {Carr}\ \emph {et~al.}(2021)\citenamefont {Carr},
  \citenamefont {Kohri}, \citenamefont {Sendouda},\ and\ \citenamefont
  {Yokoyama}}]{Carr:2020gox}%
  \BibitemOpen
  \bibfield  {author} {\bibinfo {author} {\bibfnamefont {B.}~\bibnamefont
  {Carr}}, \bibinfo {author} {\bibfnamefont {K.}~\bibnamefont {Kohri}},
  \bibinfo {author} {\bibfnamefont {Y.}~\bibnamefont {Sendouda}}, \ and\
  \bibinfo {author} {\bibfnamefont {J.}~\bibnamefont {Yokoyama}},\ }\href
  {\doibase 10.1088/1361-6633/ac1e31} {\bibfield  {journal} {\bibinfo
  {journal} {Rept. Prog. Phys.}\ }\textbf {\bibinfo {volume} {84}},\ \bibinfo
  {pages} {116902} (\bibinfo {year} {2021})},\ \Eprint
  {http://arxiv.org/abs/2002.12778} {arXiv:2002.12778 [astro-ph.CO]}
  \BibitemShut {NoStop}%
\bibitem [{\citenamefont {Inomata}\ \emph {et~al.}(2020)\citenamefont
  {Inomata}, \citenamefont {Kawasaki}, \citenamefont {Mukaida}, \citenamefont
  {Terada},\ and\ \citenamefont {Yanagida}}]{Inomata:2020lmk}%
  \BibitemOpen
  \bibfield  {author} {\bibinfo {author} {\bibfnamefont {K.}~\bibnamefont
  {Inomata}}, \bibinfo {author} {\bibfnamefont {M.}~\bibnamefont {Kawasaki}},
  \bibinfo {author} {\bibfnamefont {K.}~\bibnamefont {Mukaida}}, \bibinfo
  {author} {\bibfnamefont {T.}~\bibnamefont {Terada}}, \ and\ \bibinfo {author}
  {\bibfnamefont {T.~T.}\ \bibnamefont {Yanagida}},\ }\href {\doibase
  10.1103/PhysRevD.101.123533} {\bibfield  {journal} {\bibinfo  {journal}
  {Phys. Rev. D}\ }\textbf {\bibinfo {volume} {101}},\ \bibinfo {pages}
  {123533} (\bibinfo {year} {2020})},\ \Eprint
  {http://arxiv.org/abs/2003.10455} {arXiv:2003.10455 [astro-ph.CO]}
  \BibitemShut {NoStop}%
\bibitem [{\citenamefont {Barman}\ \emph {et~al.}(2022)\citenamefont {Barman},
  \citenamefont {Borah}, \citenamefont {Jyoti~Das},\ and\ \citenamefont
  {Roshan}}]{Barman:2022pdo}%
  \BibitemOpen
  \bibfield  {author} {\bibinfo {author} {\bibfnamefont {B.}~\bibnamefont
  {Barman}}, \bibinfo {author} {\bibfnamefont {D.}~\bibnamefont {Borah}},
  \bibinfo {author} {\bibfnamefont {S.}~\bibnamefont {Jyoti~Das}}, \ and\
  \bibinfo {author} {\bibfnamefont {R.}~\bibnamefont {Roshan}},\ }\href@noop {}
  {\  (\bibinfo {year} {2022})},\ \Eprint {http://arxiv.org/abs/2212.00052}
  {arXiv:2212.00052 [hep-ph]} \BibitemShut {NoStop}%
\bibitem [{\citenamefont {Bhaumik}\ \emph {et~al.}(2022)\citenamefont
  {Bhaumik}, \citenamefont {Ghoshal}, \citenamefont {Jain},\ and\ \citenamefont
  {Lewicki}}]{Bhaumik:2022zdd}%
  \BibitemOpen
  \bibfield  {author} {\bibinfo {author} {\bibfnamefont {N.}~\bibnamefont
  {Bhaumik}}, \bibinfo {author} {\bibfnamefont {A.}~\bibnamefont {Ghoshal}},
  \bibinfo {author} {\bibfnamefont {R.~K.}\ \bibnamefont {Jain}}, \ and\
  \bibinfo {author} {\bibfnamefont {M.}~\bibnamefont {Lewicki}},\ }\href@noop
  {} {\  (\bibinfo {year} {2022})},\ \Eprint {http://arxiv.org/abs/2212.00775}
  {arXiv:2212.00775 [astro-ph.CO]} \BibitemShut {NoStop}%
\bibitem [{\citenamefont {Morrison}\ \emph {et~al.}(2019)\citenamefont
  {Morrison}, \citenamefont {Profumo},\ and\ \citenamefont
  {Yu}}]{Morrison:2018xla}%
  \BibitemOpen
  \bibfield  {author} {\bibinfo {author} {\bibfnamefont {L.}~\bibnamefont
  {Morrison}}, \bibinfo {author} {\bibfnamefont {S.}~\bibnamefont {Profumo}}, \
  and\ \bibinfo {author} {\bibfnamefont {Y.}~\bibnamefont {Yu}},\ }\href
  {\doibase 10.1088/1475-7516/2019/05/005} {\bibfield  {journal} {\bibinfo
  {journal} {JCAP}\ }\textbf {\bibinfo {volume} {1905}},\ \bibinfo {pages}
  {005} (\bibinfo {year} {2019})},\ \Eprint {http://arxiv.org/abs/1812.10606}
  {arXiv:1812.10606 [astro-ph.CO]} \BibitemShut {NoStop}%
\bibitem [{\citenamefont {Gondolo}\ \emph {et~al.}(2020)\citenamefont
  {Gondolo}, \citenamefont {Sandick},\ and\ \citenamefont {Shams
  Es~Haghi}}]{Gondolo:2020uqv}%
  \BibitemOpen
  \bibfield  {author} {\bibinfo {author} {\bibfnamefont {P.}~\bibnamefont
  {Gondolo}}, \bibinfo {author} {\bibfnamefont {P.}~\bibnamefont {Sandick}}, \
  and\ \bibinfo {author} {\bibfnamefont {B.}~\bibnamefont {Shams Es~Haghi}},\
  }\href {\doibase 10.1103/PhysRevD.102.095018} {\bibfield  {journal} {\bibinfo
   {journal} {Phys. Rev. D}\ }\textbf {\bibinfo {volume} {102}},\ \bibinfo
  {pages} {095018} (\bibinfo {year} {2020})},\ \Eprint
  {http://arxiv.org/abs/2009.02424} {arXiv:2009.02424 [hep-ph]} \BibitemShut
  {NoStop}%
\bibitem [{\citenamefont {Bernal}\ and\ \citenamefont
  {Zapata}(2021{\natexlab{a}})}]{Bernal:2020ili}%
  \BibitemOpen
  \bibfield  {author} {\bibinfo {author} {\bibfnamefont {N.}~\bibnamefont
  {Bernal}}\ and\ \bibinfo {author} {\bibfnamefont {O.}~\bibnamefont
  {Zapata}},\ }\href {\doibase 10.1016/j.physletb.2021.136129} {\bibfield
  {journal} {\bibinfo  {journal} {Phys. Lett. B}\ }\textbf {\bibinfo {volume}
  {815}},\ \bibinfo {pages} {136129} (\bibinfo {year} {2021}{\natexlab{a}})},\
  \Eprint {http://arxiv.org/abs/2011.02510} {arXiv:2011.02510 [hep-ph]}
  \BibitemShut {NoStop}%
\bibitem [{\citenamefont {Jyoti~Das}\ \emph {et~al.}(2021)\citenamefont
  {Jyoti~Das}, \citenamefont {Mahanta},\ and\ \citenamefont
  {Borah}}]{JyotiDas:2021shi}%
  \BibitemOpen
  \bibfield  {author} {\bibinfo {author} {\bibfnamefont {S.}~\bibnamefont
  {Jyoti~Das}}, \bibinfo {author} {\bibfnamefont {D.}~\bibnamefont {Mahanta}},
  \ and\ \bibinfo {author} {\bibfnamefont {D.}~\bibnamefont {Borah}},\
  }\href@noop {} {\  (\bibinfo {year} {2021})},\ \Eprint
  {http://arxiv.org/abs/2104.14496} {arXiv:2104.14496 [hep-ph]} \BibitemShut
  {NoStop}%
\bibitem [{\citenamefont {Hamada}\ and\ \citenamefont
  {Iso}(2017)}]{Hamada:2016jnq}%
  \BibitemOpen
  \bibfield  {author} {\bibinfo {author} {\bibfnamefont {Y.}~\bibnamefont
  {Hamada}}\ and\ \bibinfo {author} {\bibfnamefont {S.}~\bibnamefont {Iso}},\
  }\href {\doibase 10.1093/ptep/ptx011} {\bibfield  {journal} {\bibinfo
  {journal} {PTEP}\ }\textbf {\bibinfo {volume} {2017}},\ \bibinfo {pages}
  {033B02} (\bibinfo {year} {2017})},\ \Eprint
  {http://arxiv.org/abs/1610.02586} {arXiv:1610.02586 [hep-ph]} \BibitemShut
  {NoStop}%
\bibitem [{\citenamefont {Hooper}\ and\ \citenamefont
  {Krnjaic}(2021)}]{Hooper:2020otu}%
  \BibitemOpen
  \bibfield  {author} {\bibinfo {author} {\bibfnamefont {D.}~\bibnamefont
  {Hooper}}\ and\ \bibinfo {author} {\bibfnamefont {G.}~\bibnamefont
  {Krnjaic}},\ }\href {\doibase 10.1103/PhysRevD.103.043504} {\bibfield
  {journal} {\bibinfo  {journal} {Phys. Rev. D}\ }\textbf {\bibinfo {volume}
  {103}},\ \bibinfo {pages} {043504} (\bibinfo {year} {2021})},\ \Eprint
  {http://arxiv.org/abs/2010.01134} {arXiv:2010.01134 [hep-ph]} \BibitemShut
  {NoStop}%
\bibitem [{\citenamefont {Chaudhuri}\ and\ \citenamefont
  {Dolgov}(2020)}]{Chaudhuri:2020wjo}%
  \BibitemOpen
  \bibfield  {author} {\bibinfo {author} {\bibfnamefont {A.}~\bibnamefont
  {Chaudhuri}}\ and\ \bibinfo {author} {\bibfnamefont {A.}~\bibnamefont
  {Dolgov}},\ }\href@noop {} {\  (\bibinfo {year} {2020})},\ \Eprint
  {http://arxiv.org/abs/2001.11219} {arXiv:2001.11219 [astro-ph.CO]}
  \BibitemShut {NoStop}%
\bibitem [{\citenamefont {Kitabayashi}(2021)}]{Kitabayashi:2021hox}%
  \BibitemOpen
  \bibfield  {author} {\bibinfo {author} {\bibfnamefont {T.}~\bibnamefont
  {Kitabayashi}},\ }\href@noop {} {\  (\bibinfo {year} {2021})},\ \Eprint
  {http://arxiv.org/abs/2101.01921} {arXiv:2101.01921 [hep-ph]} \BibitemShut
  {NoStop}%
\bibitem [{\citenamefont {Cheek}\ \emph
  {et~al.}(2022{\natexlab{a}})\citenamefont {Cheek}, \citenamefont {Heurtier},
  \citenamefont {Perez-Gonzalez},\ and\ \citenamefont
  {Turner}}]{Cheek:2021odj}%
  \BibitemOpen
  \bibfield  {author} {\bibinfo {author} {\bibfnamefont {A.}~\bibnamefont
  {Cheek}}, \bibinfo {author} {\bibfnamefont {L.}~\bibnamefont {Heurtier}},
  \bibinfo {author} {\bibfnamefont {Y.~F.}\ \bibnamefont {Perez-Gonzalez}}, \
  and\ \bibinfo {author} {\bibfnamefont {J.}~\bibnamefont {Turner}},\ }\href
  {\doibase 10.1103/PhysRevD.105.015022} {\bibfield  {journal} {\bibinfo
  {journal} {Phys. Rev. D}\ }\textbf {\bibinfo {volume} {105}},\ \bibinfo
  {pages} {015022} (\bibinfo {year} {2022}{\natexlab{a}})},\ \Eprint
  {http://arxiv.org/abs/2107.00013} {arXiv:2107.00013 [hep-ph]} \BibitemShut
  {NoStop}%
\bibitem [{\citenamefont {Martin}\ \emph
  {et~al.}(2020{\natexlab{a}})\citenamefont {Martin}, \citenamefont
  {Papanikolaou},\ and\ \citenamefont {Vennin}}]{Martin:2019nuw}%
  \BibitemOpen
  \bibfield  {author} {\bibinfo {author} {\bibfnamefont {J.}~\bibnamefont
  {Martin}}, \bibinfo {author} {\bibfnamefont {T.}~\bibnamefont
  {Papanikolaou}}, \ and\ \bibinfo {author} {\bibfnamefont {V.}~\bibnamefont
  {Vennin}},\ }\href {\doibase 10.1088/1475-7516/2020/01/024} {\bibfield
  {journal} {\bibinfo  {journal} {JCAP}\ }\textbf {\bibinfo {volume} {01}},\
  \bibinfo {pages} {024} (\bibinfo {year} {2020}{\natexlab{a}})},\ \Eprint
  {http://arxiv.org/abs/1907.04236} {arXiv:1907.04236 [astro-ph.CO]}
  \BibitemShut {NoStop}%
\bibitem [{\citenamefont {Martin}\ \emph
  {et~al.}(2020{\natexlab{b}})\citenamefont {Martin}, \citenamefont
  {Papanikolaou}, \citenamefont {Pinol},\ and\ \citenamefont
  {Vennin}}]{Martin:2020fgl}%
  \BibitemOpen
  \bibfield  {author} {\bibinfo {author} {\bibfnamefont {J.}~\bibnamefont
  {Martin}}, \bibinfo {author} {\bibfnamefont {T.}~\bibnamefont
  {Papanikolaou}}, \bibinfo {author} {\bibfnamefont {L.}~\bibnamefont {Pinol}},
  \ and\ \bibinfo {author} {\bibfnamefont {V.}~\bibnamefont {Vennin}},\ }\href
  {\doibase 10.1088/1475-7516/2020/05/003} {\bibfield  {journal} {\bibinfo
  {journal} {JCAP}\ }\textbf {\bibinfo {volume} {05}},\ \bibinfo {pages} {003}
  (\bibinfo {year} {2020}{\natexlab{b}})},\ \Eprint
  {http://arxiv.org/abs/2002.01820} {arXiv:2002.01820 [astro-ph.CO]}
  \BibitemShut {NoStop}%
\bibitem [{\citenamefont {Auclair}\ and\ \citenamefont
  {Vennin}(2021)}]{Auclair:2020csm}%
  \BibitemOpen
  \bibfield  {author} {\bibinfo {author} {\bibfnamefont {P.}~\bibnamefont
  {Auclair}}\ and\ \bibinfo {author} {\bibfnamefont {V.}~\bibnamefont
  {Vennin}},\ }\href {\doibase 10.1088/1475-7516/2021/02/038} {\bibfield
  {journal} {\bibinfo  {journal} {JCAP}\ }\textbf {\bibinfo {volume} {02}},\
  \bibinfo {pages} {038} (\bibinfo {year} {2021})},\ \Eprint
  {http://arxiv.org/abs/2011.05633} {arXiv:2011.05633 [astro-ph.CO]}
  \BibitemShut {NoStop}%
\bibitem [{\citenamefont {Cheek}\ \emph
  {et~al.}(2022{\natexlab{b}})\citenamefont {Cheek}, \citenamefont {Heurtier},
  \citenamefont {Perez-Gonzalez},\ and\ \citenamefont
  {Turner}}]{Cheek:2022dbx}%
  \BibitemOpen
  \bibfield  {author} {\bibinfo {author} {\bibfnamefont {A.}~\bibnamefont
  {Cheek}}, \bibinfo {author} {\bibfnamefont {L.}~\bibnamefont {Heurtier}},
  \bibinfo {author} {\bibfnamefont {Y.~F.}\ \bibnamefont {Perez-Gonzalez}}, \
  and\ \bibinfo {author} {\bibfnamefont {J.}~\bibnamefont {Turner}},\
  }\href@noop {} {\  (\bibinfo {year} {2022}{\natexlab{b}})},\ \Eprint
  {http://arxiv.org/abs/2207.09462} {arXiv:2207.09462 [astro-ph.CO]}
  \BibitemShut {NoStop}%
\bibitem [{\citenamefont {Hooper}\ \emph {et~al.}(2019)\citenamefont {Hooper},
  \citenamefont {Krnjaic},\ and\ \citenamefont {McDermott}}]{Hooper:2019gtx}%
  \BibitemOpen
  \bibfield  {author} {\bibinfo {author} {\bibfnamefont {D.}~\bibnamefont
  {Hooper}}, \bibinfo {author} {\bibfnamefont {G.}~\bibnamefont {Krnjaic}}, \
  and\ \bibinfo {author} {\bibfnamefont {S.~D.}\ \bibnamefont {McDermott}},\
  }\href {\doibase 10.1007/JHEP08(2019)001} {\bibfield  {journal} {\bibinfo
  {journal} {JHEP}\ }\textbf {\bibinfo {volume} {08}},\ \bibinfo {pages} {001}
  (\bibinfo {year} {2019})},\ \Eprint {http://arxiv.org/abs/1905.01301}
  {arXiv:1905.01301 [hep-ph]} \BibitemShut {NoStop}%
\bibitem [{\citenamefont {Cheek}\ \emph
  {et~al.}(2022{\natexlab{c}})\citenamefont {Cheek}, \citenamefont {Heurtier},
  \citenamefont {Perez-Gonzalez},\ and\ \citenamefont
  {Turner}}]{Cheek:2021cfe}%
  \BibitemOpen
  \bibfield  {author} {\bibinfo {author} {\bibfnamefont {A.}~\bibnamefont
  {Cheek}}, \bibinfo {author} {\bibfnamefont {L.}~\bibnamefont {Heurtier}},
  \bibinfo {author} {\bibfnamefont {Y.~F.}\ \bibnamefont {Perez-Gonzalez}}, \
  and\ \bibinfo {author} {\bibfnamefont {J.}~\bibnamefont {Turner}},\ }\href
  {\doibase 10.1103/PhysRevD.105.015023} {\bibfield  {journal} {\bibinfo
  {journal} {Phys. Rev. D}\ }\textbf {\bibinfo {volume} {105}},\ \bibinfo
  {pages} {015023} (\bibinfo {year} {2022}{\natexlab{c}})},\ \Eprint
  {http://arxiv.org/abs/2107.00016} {arXiv:2107.00016 [hep-ph]} \BibitemShut
  {NoStop}%
\bibitem [{\citenamefont {Perez-Gonzalez}\ and\ \citenamefont
  {Turner}(2020)}]{Perez-Gonzalez:2020vnz}%
  \BibitemOpen
  \bibfield  {author} {\bibinfo {author} {\bibfnamefont {Y.~F.}\ \bibnamefont
  {Perez-Gonzalez}}\ and\ \bibinfo {author} {\bibfnamefont {J.}~\bibnamefont
  {Turner}},\ }\href@noop {} {\  (\bibinfo {year} {2020})},\ \Eprint
  {http://arxiv.org/abs/2010.03565} {arXiv:2010.03565 [hep-ph]} \BibitemShut
  {NoStop}%
\bibitem [{\citenamefont {Bernal}\ and\ \citenamefont
  {Zapata}(2021{\natexlab{b}})}]{Bernal:2020kse}%
  \BibitemOpen
  \bibfield  {author} {\bibinfo {author} {\bibfnamefont {N.}~\bibnamefont
  {Bernal}}\ and\ \bibinfo {author} {\bibfnamefont {O.}~\bibnamefont
  {Zapata}},\ }\href {\doibase 10.1088/1475-7516/2021/03/007} {\bibfield
  {journal} {\bibinfo  {journal} {JCAP}\ }\textbf {\bibinfo {volume} {03}},\
  \bibinfo {pages} {007} (\bibinfo {year} {2021}{\natexlab{b}})},\ \Eprint
  {http://arxiv.org/abs/2010.09725} {arXiv:2010.09725 [hep-ph]} \BibitemShut
  {NoStop}%
\bibitem [{\citenamefont {Datta}\ \emph {et~al.}(2020)\citenamefont {Datta},
  \citenamefont {Ghosal},\ and\ \citenamefont {Samanta}}]{Datta:2020bht}%
  \BibitemOpen
  \bibfield  {author} {\bibinfo {author} {\bibfnamefont {S.}~\bibnamefont
  {Datta}}, \bibinfo {author} {\bibfnamefont {A.}~\bibnamefont {Ghosal}}, \
  and\ \bibinfo {author} {\bibfnamefont {R.}~\bibnamefont {Samanta}},\
  }\href@noop {} {\  (\bibinfo {year} {2020})},\ \Eprint
  {http://arxiv.org/abs/2012.14981} {arXiv:2012.14981 [hep-ph]} \BibitemShut
  {NoStop}%
\bibitem [{\citenamefont {Masina}(2021)}]{Masina:2021zpu}%
  \BibitemOpen
  \bibfield  {author} {\bibinfo {author} {\bibfnamefont {I.}~\bibnamefont
  {Masina}},\ }\href@noop {} {\  (\bibinfo {year} {2021})},\ \Eprint
  {http://arxiv.org/abs/2103.13825} {arXiv:2103.13825 [gr-qc]} \BibitemShut
  {NoStop}%
\bibitem [{\citenamefont {Bernal}\ and\ \citenamefont
  {Zapata}(2021{\natexlab{c}})}]{Bernal:2020bjf}%
  \BibitemOpen
  \bibfield  {author} {\bibinfo {author} {\bibfnamefont {N.}~\bibnamefont
  {Bernal}}\ and\ \bibinfo {author} {\bibfnamefont {O.}~\bibnamefont
  {Zapata}},\ }\href {\doibase 10.1088/1475-7516/2021/03/015} {\bibfield
  {journal} {\bibinfo  {journal} {JCAP}\ }\textbf {\bibinfo {volume} {03}},\
  \bibinfo {pages} {015} (\bibinfo {year} {2021}{\natexlab{c}})},\ \Eprint
  {http://arxiv.org/abs/2011.12306} {arXiv:2011.12306 [astro-ph.CO]}
  \BibitemShut {NoStop}%
\bibitem [{\citenamefont {Baldes}\ \emph {et~al.}(2020)\citenamefont {Baldes},
  \citenamefont {Decant}, \citenamefont {Hooper},\ and\ \citenamefont
  {Lopez-Honorez}}]{Baldes:2020nuv}%
  \BibitemOpen
  \bibfield  {author} {\bibinfo {author} {\bibfnamefont {I.}~\bibnamefont
  {Baldes}}, \bibinfo {author} {\bibfnamefont {Q.}~\bibnamefont {Decant}},
  \bibinfo {author} {\bibfnamefont {D.~C.}\ \bibnamefont {Hooper}}, \ and\
  \bibinfo {author} {\bibfnamefont {L.}~\bibnamefont {Lopez-Honorez}},\ }\href
  {\doibase 10.1088/1475-7516/2020/08/045} {\bibfield  {journal} {\bibinfo
  {journal} {JCAP}\ }\textbf {\bibinfo {volume} {08}},\ \bibinfo {pages} {045}
  (\bibinfo {year} {2020})},\ \Eprint {http://arxiv.org/abs/2004.14773}
  {arXiv:2004.14773 [astro-ph.CO]} \BibitemShut {NoStop}%
\bibitem [{\citenamefont {Allahverdi}\ \emph {et~al.}(2018)\citenamefont
  {Allahverdi}, \citenamefont {Dent},\ and\ \citenamefont
  {Osinski}}]{Allahverdi:2017sks}%
  \BibitemOpen
  \bibfield  {author} {\bibinfo {author} {\bibfnamefont {R.}~\bibnamefont
  {Allahverdi}}, \bibinfo {author} {\bibfnamefont {J.}~\bibnamefont {Dent}}, \
  and\ \bibinfo {author} {\bibfnamefont {J.}~\bibnamefont {Osinski}},\ }\href
  {\doibase 10.1103/PhysRevD.97.055013} {\bibfield  {journal} {\bibinfo
  {journal} {Phys. Rev.}\ }\textbf {\bibinfo {volume} {D97}},\ \bibinfo {pages}
  {055013} (\bibinfo {year} {2018})},\ \Eprint
  {http://arxiv.org/abs/1711.10511} {arXiv:1711.10511 [astro-ph.CO]}
  \BibitemShut {NoStop}%
\bibitem [{\citenamefont {Masina}(2020)}]{Masina:2020xhk}%
  \BibitemOpen
  \bibfield  {author} {\bibinfo {author} {\bibfnamefont {I.}~\bibnamefont
  {Masina}},\ }\href {\doibase 10.1140/epjp/s13360-020-00564-9} {\bibfield
  {journal} {\bibinfo  {journal} {Eur. Phys. J. Plus}\ }\textbf {\bibinfo
  {volume} {135}},\ \bibinfo {pages} {552} (\bibinfo {year} {2020})},\ \Eprint
  {http://arxiv.org/abs/2004.04740} {arXiv:2004.04740 [hep-ph]} \BibitemShut
  {NoStop}%
\bibitem [{\citenamefont {Lunardini}\ and\ \citenamefont
  {Perez-Gonzalez}(2020)}]{Lunardini:2019zob}%
  \BibitemOpen
  \bibfield  {author} {\bibinfo {author} {\bibfnamefont {C.}~\bibnamefont
  {Lunardini}}\ and\ \bibinfo {author} {\bibfnamefont {Y.~F.}\ \bibnamefont
  {Perez-Gonzalez}},\ }\href {\doibase 10.1088/1475-7516/2020/08/014}
  {\bibfield  {journal} {\bibinfo  {journal} {JCAP}\ }\textbf {\bibinfo
  {volume} {08}},\ \bibinfo {pages} {014} (\bibinfo {year} {2020})},\ \Eprint
  {http://arxiv.org/abs/1910.07864} {arXiv:1910.07864 [hep-ph]} \BibitemShut
  {NoStop}%
\bibitem [{\citenamefont {Dienes}\ \emph {et~al.}(2022)\citenamefont {Dienes},
  \citenamefont {Heurtier}, \citenamefont {Huang}, \citenamefont {Kim},
  \citenamefont {Tait},\ and\ \citenamefont {Thomas}}]{Dienes:2022zgd}%
  \BibitemOpen
  \bibfield  {author} {\bibinfo {author} {\bibfnamefont {K.~R.}\ \bibnamefont
  {Dienes}}, \bibinfo {author} {\bibfnamefont {L.}~\bibnamefont {Heurtier}},
  \bibinfo {author} {\bibfnamefont {F.}~\bibnamefont {Huang}}, \bibinfo
  {author} {\bibfnamefont {D.}~\bibnamefont {Kim}}, \bibinfo {author}
  {\bibfnamefont {T.~M.~P.}\ \bibnamefont {Tait}}, \ and\ \bibinfo {author}
  {\bibfnamefont {B.}~\bibnamefont {Thomas}},\ }\href@noop {} {\  (\bibinfo
  {year} {2022})},\ \Eprint {http://arxiv.org/abs/2212.01369} {arXiv:2212.01369
  [astro-ph.CO]} \BibitemShut {NoStop}%
\bibitem [{\citenamefont {Papanikolaou}(2022)}]{Papanikolaou:2022chm}%
  \BibitemOpen
  \bibfield  {author} {\bibinfo {author} {\bibfnamefont {T.}~\bibnamefont
  {Papanikolaou}},\ }\href {\doibase 10.1088/1475-7516/2022/10/089} {\bibfield
  {journal} {\bibinfo  {journal} {JCAP}\ }\textbf {\bibinfo {volume} {10}},\
  \bibinfo {pages} {089} (\bibinfo {year} {2022})},\ \Eprint
  {http://arxiv.org/abs/2207.11041} {arXiv:2207.11041 [astro-ph.CO]}
  \BibitemShut {NoStop}%
\bibitem [{\citenamefont {Saito}\ and\ \citenamefont
  {Yokoyama}(2009)}]{Saito:2008jc}%
  \BibitemOpen
  \bibfield  {author} {\bibinfo {author} {\bibfnamefont {R.}~\bibnamefont
  {Saito}}\ and\ \bibinfo {author} {\bibfnamefont {J.}~\bibnamefont
  {Yokoyama}},\ }\href {\doibase 10.1103/PhysRevLett.102.161101} {\bibfield
  {journal} {\bibinfo  {journal} {Phys. Rev. Lett.}\ }\textbf {\bibinfo
  {volume} {102}},\ \bibinfo {pages} {161101} (\bibinfo {year} {2009})},\
  \bibinfo {note} {[Erratum: Phys.Rev.Lett. 107, 069901 (2011)]},\ \Eprint
  {http://arxiv.org/abs/0812.4339} {arXiv:0812.4339 [astro-ph]} \BibitemShut
  {NoStop}%
\end{thebibliography}%

\end{document}